\documentclass[10pt, conference, compsocconf]{IEEEtran}
\usepackage{graphicx}
\usepackage{paralist}
\usepackage[lined]{algorithm2e}
\usepackage{float}
\usepackage{epstopdf}

\newfloat{algorithm}{t}{lop}

\ifCLASSINFOpdf
\else
\fi
\usepackage[cmex10]{amsmath}
\usepackage{amsfonts}

\begin{document}
%
\title{Content as a Network Primitive}


\author{
\IEEEauthorblockN{Abhishek Chanda}
\IEEEauthorblockA{WINLAB, Rutgers University\\
North Brunswick, NJ, USA\\
achanda@winlab.rutgers.edu}
\and
\IEEEauthorblockN{Cedric Westphal}
\IEEEauthorblockA{Innovation Center\\
Huawei Technology\\
Santa Clara, CA, USA\\
cwestphal@huawei.com}
}

\maketitle

\begin{abstract}
The current functionality supported by OpenFlow-based software defined networking (SDN) includes switching, routing, tunneling, and some basic fire walling while operating on traffic flows. However, the semantics of SDN do not allow for other operations on the traffic, nor does it allow operations at a higher granularity. In this work, we describe a method to expand the SDN framework to add other network primitives. In particular, we present a method to integrate different network elements (like cache, proxy etc). Here, we focus on storage and caching, but our method could be expanded to other functionality seamlessly. We also present a method to identify content so as to perform per-content policy, as opposed to per flow policy. We have implemented the proposed mechanisms to demonstrate its feasibility.

\end{abstract}

\begin{IEEEkeywords}
SDN, content management, network abstractions, 	Cache storage

\end{IEEEkeywords}

%
\IEEEpeerreviewmaketitle

\section{Introduction}

Software defined networking (SDN) decouples the control plane and the forwarding plane of a network. The forwarding plane then exports some simple APIs to the control plane, which then utilizes these APIs to provide desired network policy. The decoupling allows for the control plane to be written in software, and thus be programmable.

Current approaches for SDN, such as OpenFlow~\cite{McKeown2008OpenFlow}, focus on controlling switching element, and adopt a definition of the forwarding plane which takes {\em traffic flows} as the unit to which a policy is applied. However, this approach suffers from two limitations:
\begin{itemize}
\item The network might include other non-forwarding elements; it is the common view of most future Internet architectures~\cite{NDN,Anand2011XIA,Paul2008Cacheandforward} and many commercial products~\cite{CiscoSRE} combine a switching element with additional capability, such as storage and processing; these capabilities need to be advertised to the control plane so that the programmable policy takes them into account.
\item The network might need to consider different policy for different objects at a finer granularity. Currently, OpenFlow ignores the specific content and provides a forwarding behavior that is per-flow based, where the flow is defined as a combination of the source and destination MAC and IP addresses and protocol ports plus some other fields such as MPLS tags, VLAN tags; however, we can consider a specific content file: while the source and destination and protocols of two files might be identical (say, from the same host to the same data center using TCP port 80), the two files might require different forwarding treatment (say, one is a streaming video with some delay constraints while the other is a picture with less strict delay requirements). Many architectures attempt to place content as the focus of the architecture~\cite{Jacobson2009Networking,Ghodsi2011InformationCentric,Koponen2007Dataoriented,Pursuit} and this is not supported by the current SDN proposals.
\end{itemize}

In this paper, we propose an extension to the current SDN model to include solutions to the two limitations identified above: namely, an extension of OpenFlow to network elements beyond switching and forwarding; and an extension of OpenFlow to handle the granularity of the content as well as the granularity of the flow. Other proposals to extend the SDN framework include integration with the application layer transport optimization~\cite{Xie2012Use}.

In this first step towards a more general definition of SDN, we focus on a particular use case. Namely, we will add {\em storage} primitives to SDN as an example of a new element that can be under the controller's reach. Other processing elements could be included as well. And we will add {\em content handling mechanisms} to enable seamless switching based on content. To demonstrate the feasibility of our approach, we have implemented a content management framework on top of a SDN controller and switches which includes a {\em distributed, transparent caching mechanism}.

Current SDNs are built upon IP-based Internet architecture. This means that creating content primitives will run into difficulties as this legacy architecture is not perfectly suited for content routing (as the efforts to create new content-oriented architecture~\cite{NDN,Anand2011XIA,Paul2008Cacheandforward} underline). Our contribution is to identify these challenges and offer some solutions that we have implemented. As our intent is to show the feasibility of such an architecture, we try to re-use existing mechanisms as often as possible.

The rest of the paper is organized as follows: Section~\ref{sec:stor} describes how to define storage primitives to register storage capability with the controller and allow the controller to direct flows to be stored in addition to route the flows through the network. Section~\ref{sec:cont} will describe our mechanism to identify content and to provide differentiated treatment to different traffic despite their matching in terms of source, destination, protocols and potentially other filtering fields. Section~\ref{sec:architecture} will show how to put these together to present an illustration of the capability enabled by our framework. Namely, we will construct a programmable content management module in the network controller in order to set up a transparent CDN. We describe in Section~\ref{sec:eval} some details of our implementation and provide some evaluation results. Section~\ref{sec:con} concludes.

\section{Storage Primitives}
\label{sec:stor}

In-network storage is expected to be deployed in most routers in many future network architectures. In a SDN architecture with an abstracted view of the network being held at a controller, in-network storage elements (storage enabled routers, cache etc.) need to be able to advertise to the controller their ability to store content. And since the controller has a high level view of the whole network, it can use the storage elements based upon its need and its network performance targets. There are two types of in-network storage:
\begin{inparaenum}
\item Forwarding elements in the network can have storage associated with them. In a first step, storage in a routing element should be understood as a store-and-forward capacity, rather than a distributed file system: namely a file can be stored in the network for traffic engineering purpose or for caching. This is for administrative reasons: we assume that the network operators own its network elements and use their resource to optimize their traffic delivery. Later on, storage can be extended to include repositories for the output of processing tasks, assuming a framework is in place to allocate and virtualize storage to a variety of entities.
\item A network can have non-forwarding elements which can store content at different levels. Again, these elements do not form a distributed file system, rather they are a part of a distributed caching system which is synchronized over some caching protocol. Currently, the most commonly used caching protocols are ICP and HTCP.
\end{inparaenum}
The current design of the Internet caching protocols do not allow virtualization of the storage elements. In this work, we propose a method to virtualize in-network storage at the controller.

Upon adding a storage element in the network, the following configuration needs to take place:
\begin{itemize}
\item The storage element needs to identify the location of the controller;
\item The storage element needs to set up a session with the controller as any other switching element (via SSH);
\item The storage element needs to advertise its capability to the controller, in a way that the controller can understand;
\item The controller needs to integrate the capability of the storage element in its control policy;
\item The storage element needs to be able to provide some usage statistics to the controller, either at periodic interval or upon request from the controller;
\item The storage element needs to be able to refresh the association with the controller by maintaining some keep-alive mechanism; and to be able to tear down the session when it terminates.
\end{itemize}

It can be observed that this list of functionality is very similar to what a forwarding element does when it is deployed in a network. Thus, we hypothesize that extending the same (or similar) set of functionality to a non-forwarding element would be possible. Essentially, we need to expand the OpenFlow protocol with some new fields to support the tasks mentioned.

\section{Content Primitives}
\label{sec:cont}

The next important step after establishing a set of storage primitives in a SDN is to establish a set of content primitives. This means that the content needs to be properly managed and identified throughout the network. We focus here on content described by HTTP requests, as it constitutes a large part of the Internet traffic. Expanding these principles described here to other protocols and file transfer mechanisms will then be achieved through simple protocol parsing in a network element.

Content based routing (and forwarding) dictates that routing should be performed based on content, and not on the source and destination address obtained by the DNS resolution (unless the DNS mapping is updated dynamically and in real time). The benefits of this approach are straightforward; popular content gets higher priority and is probably cached somewhere close to where it is popular, resulting in reduced access latency. To see this in action, assume that a host would like to access file A and file B, which are both located on server S. However, file A happens to be very popular, and file B is not popular at all. This means that a copy of file A has been most likely replicated in some caches, while file B is exclusively found at its publisher's location, server S. Routing based upon the typical IP-based match filters would route both HTTP requests for file A and for file B to server S. Routing on content on the other hand would let the request for file B go to server S, and the request for file A would be send to a nearby cache which contains the requested object. And if the cache is located somewhere close to the clients who request the content, this would result in decreased latency and delay.

The following needs to be supported:
\begin{itemize}
\item Content identification: a network element must be able to identify content preferably close to the source. There can be a number of possible methods of doing this, content can be identified based on a combination of HTTP and TCP semantics (all traffic on port 80 on a webserver is content). Another option is to put a DPI module on the customer facing edge of the network. The third and the most robust option is to include special content matching fields in OpenFlow that forwarding elements can match against.
\item Content routing: once content has been identified and the destination has been found, we would need to route the packets to a given destination. This can be implemented by augmenting an OpenFlow controller with a module that can push custom flows to forwarding elements. 
\item Integration with the cache mechanism: standard Internet caching systems are complicated. They expect a set of HTTP headers to configure caching behavior. It is a design decision whether an OpenFlow based solution comply with existing Internet caching mechanisms. For the sake of brevity, in this work we ignored integration with Internet caching.
\end{itemize}

The next section describes our proposed architecture in more detail.

\section{Description of the architecture}
\label{sec:architecture}

\subsection{Overview}
\begin{figure}[h]
\centering
\includegraphics[scale=0.4]{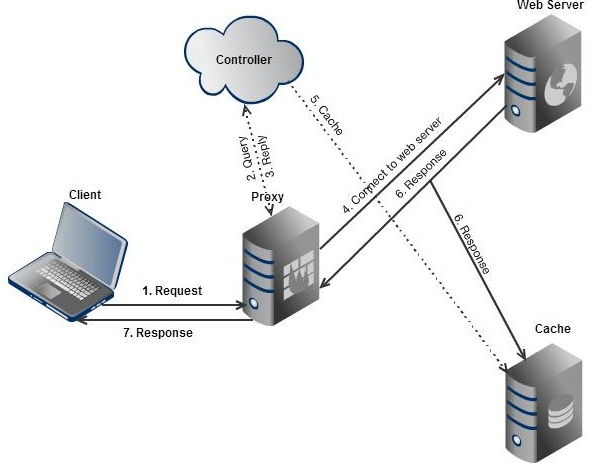}
\caption{Overview of the architecture}
\label{fig:arch}
\end{figure}

Like in most caching architectures, the major design philosophy here is \emph{separation of content and its metadata}. Thus, we try to separate out metadata from content close to the source and put it in the control plane. In our implementation, the metadata consists of the file name parsed from the HTTP GET request and the TCP flow information. Thus, it is a five tuple of the form \emph{$\langle file~ name, destination ~IP, destination~ port, source~ IP, \\source~ port \rangle$}. A high level overview of the architecture is shown in figure \ref{fig:arch}.

One key aspect is that content routing in a SDN network requires to proxy all TCP connections at the ingress switch in the network. The reason for this is that current SDN architectures still use the existing TCP semantics: a session is first established between a client and the server, before an HTTP GET is issued to request a specific content. Routing is thus performed \emph{before} the content is requested, which prohibits content routing. Proxying TCP at the ingress switch ensures that no TCP session is established beyond the ingress switch, and when the content is requested, then the proper route (and TCP session) can then be set-up.

\begin{figure*}
\centering
\includegraphics[scale=0.5]{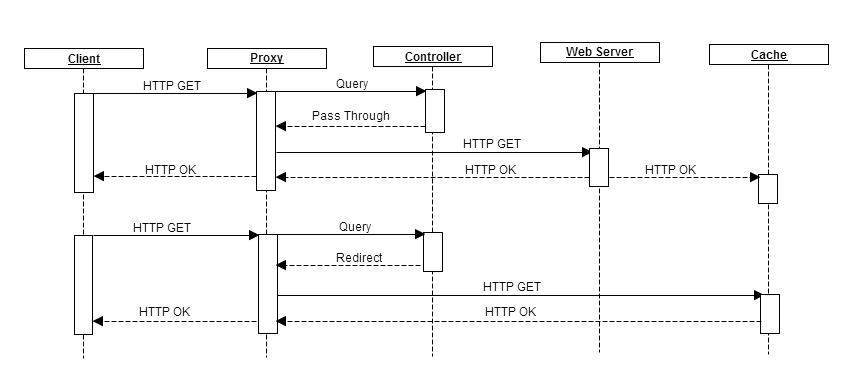}
\caption{Sequence diagram of the system}
\label{fig:sequence}
\end{figure*}

The system works as follows:
\begin{itemize}
\item The controller installs static rules into the switch directly connected to the client to forward all HTTP traffic from the client to the proxy and vice-versa. The TCP proxy terminates all TCP connections from the client.
\item The client issues an HTTP request which goes to the proxy.
\item The proxy parses the request to extract the name of the content and the web server to which the request was sent.
\item The proxy queries the controller with the file name (as a URI) asking if the file is cached somewhere in the network.
\item If the controller returns an IP and a port number, the proxy redirects the client's connection to that address. Otherwise, the proxy updates the controller with the metadata from HTTP request and directs the connection to the original web server.
\item The controller then determines whether the content from the web server should (or should not) be cached. To cache the content, it computes a forking point where the connection from the web server to the proxy should be forked so that a copy of the packets will be duplicated to the cache.
\item The controller installs a rule in the switch and invokes the cache. The controller notifies the cache of the content name and of the flow information to map the content stream to the proper content name. The controller also records the location of the cached content.
\item The cache saves the content with the file name obtained from the controller.
\end{itemize}
Figure \ref{fig:sequence} shows how the system works.

\subsection{Proxy}
The proxy is a transparent TCP proxy that is located in the client's network. In our setup, OpenFlow is used to make the proxy transparent.
The proxy is the primary device that is responsible for separating out content metadata and putting it on the control plane, thus it must intercept packets from the client and parse relevant information. The proxy communicates with the controller through REST API calls. Algorithm \ref{alg:proxy} describes how the proxy works.

\begin{algorithm}
    \SetAlgoLined
    Listen on proxy port;\\
    \eIf{a GET request arrives} {
        Parse the file name from the request;\\
        Query controller with the file name;\\
        \eIf{the controller returns an IP address} {
            Redirect all requests to that IP address;
        }
        {
            Update controller with the file name;\\
            Pass the request unmodified;
        }
        }
    {
    Do not proxy;
    }
 \caption{Proxy algorithm}
 \label{alg:proxy}
\end{algorithm}

\vspace{-5pt}

\subsection{Cache}
In our design, when the cache receives a content stream to store, the cache will see only responses from the web server. The client's (more accurately, proxy's) side of the TCP session is not redirected to the cache. This scenario is not like generic HTTP caching systems like Squid which can see both the request and response and cache the response. In our design, we want to avoid the extra round trip delay of a cache miss, so we implemented a custom cache that can accept responses and store them against request metadata obtained from the controller. The cache implements algorithm \ref{alg:cache}.

\begin{algorithm}
    \SetAlgoLined
    Listen on cache port;\\
    Start webserver on cache directory;\\
    \eIf{a HTTP response arrives} {
        Lookup the source IP of the response;\\
        Query the controller with the source IP;\\
        \eIf {the controller sends back a file name} {
            Save the response with the file name;
        }
        {
            Discard the response\;
        }
    }
    {
        Serve back the file using the webserver\;
    }
    \caption{Cache algorithm}
        \label{alg:cache}
\end{algorithm}

\vspace{-5pt}

\subsection{Controller}
The controller can run on any device that can communicate with the switch. It maintains two dictionaries that can be looked up in constant time. The $cacheDictionary$ maps file names to cache IP where the file is stored, this acts as a global dictionary for all content in a network. $requestDictionary$ maps destination server IP to file name, this is necessary to forward content mete data to the cache when it will save a content. The controller algorithm is described in \ref{alg:controller}\\

\begin{algorithm}
    \SetAlgoLined
    Install static flows for NAT in the switch to which the client is connected;\\
    $cacheDictionary \gets \{ \}$\\
    $requestDictionary \gets \{ \}$\\
    \If{proxy queries with a file name} {Lookup cache IP from $cacheDictionary$ and send back  }
    \ElseIf{proxy sends content meta data} {Insert file name and destination IP to $requestDictionary$\\
                                            Compute the forking point for a flow from destination IP to proxy and destination IP to cache\\
                                            Push flows to all switches as necessary\\
                                            Invoke the cache and send it the file name from $requestDictionary$\\
                                            Insert the file name and cache IP in $cacheDictionary$ }
    \caption{Controller algorithm}
    \label{alg:controller}
\end{algorithm}

\vspace{-5pt}

\section{Implementation and Evaluation}
\label{sec:eval}

\begin{figure*}
\centering
\includegraphics[scale=0.4]{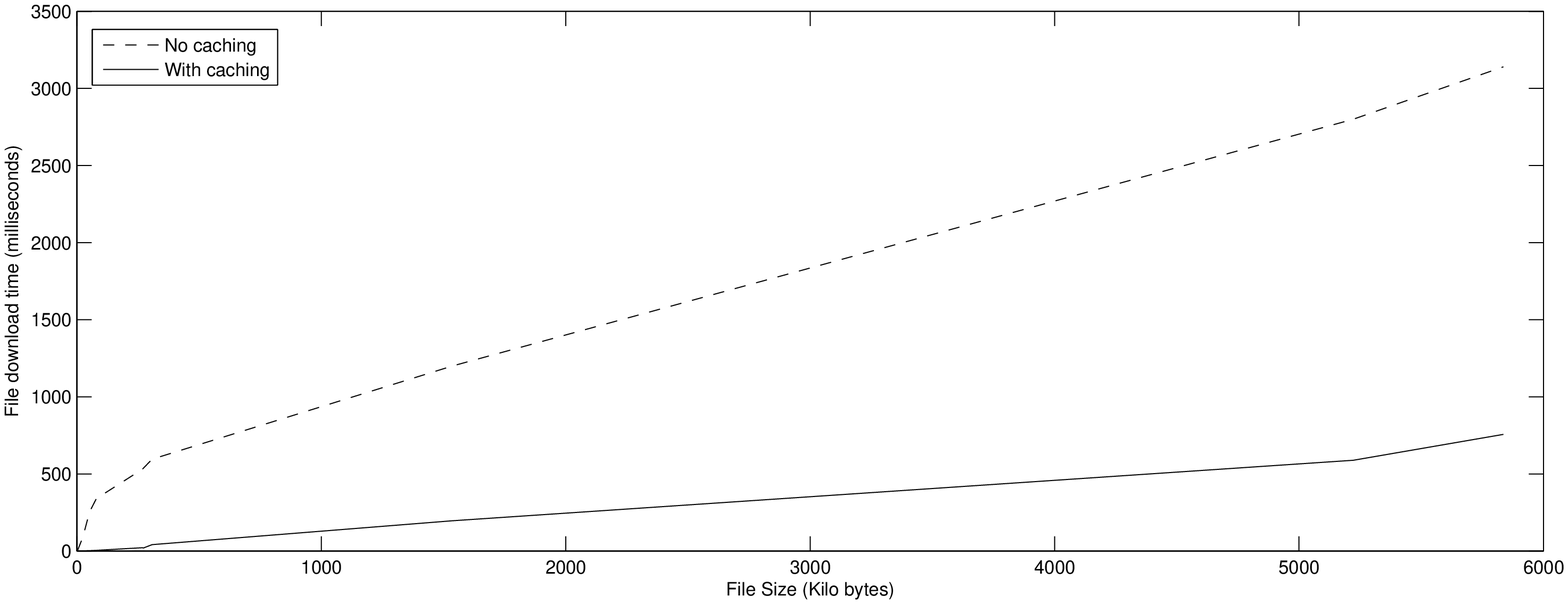}
\end{figure*}

To evaluate the merits of in-network storage virtualization, we implemented our proposed architecture on a small testbed. The testbed has a blade server (we would call it \emph{H}) running Ubuntu. The server runs an instance of Open vSwitch which enables it to act as an OpenFlow switch. \emph{H} runs three VMs each of which hosts the cache, the proxy and the client and also the FloodLight controller. This setup is placed in Santa Clara, CA.


The major components of the system are described next.
\begin{itemize}
\item We used FloodLight as the OpenFlow controller. FloodLight allows loading custom modules on the controller platform which can then write flows to all connected switches. We implemented a module to do content based forwarding on top of FloodLight.
\item We used \emph{tproxy} as the TCP proxy and wrote our own proxy script on top of it.
\item The cache is written in C++. It listens on an ethernet device on the VM and collects packets. When it sees a TCP FIN flag for a connection, it assembles data from the connection. It discards retransmitted packets and then re-arranges packets based on sequence number and writes to a file on the disk. A python script listens for file write events on the disk and when the cache saves the file, it gets triggered. It gets the file name from the controller and saves the file in the disk.
\item We placed $12$ files of different sizes, from $2Kb$ to $6Mb$ on a web server located in New Brunswick, NJ. These files are then accessed from our client in Santa Clara, CA.
\item The client access the resources in the server using a browser. We turned off browser cache for our experiments.
\end{itemize}

Now the client opens up a browser and accesses a resource in the web server. FireBug is used to measure content access delay in the two cases case, once when there is a cache miss (and the content gets stored in the cache) and the cache hit case (when the content is delivered from the cache). Caching content (and its performance benefit) is well known and we do not claim that our method innovate in this dimension. We only use this to demonstrate that our architecture works in offering content APIs to a controller and that we have successfully extended the SDN architecture to support content-based management and information-centric network ability.

\section{Related Work}

While distributed caching and software defined networking has been explored as separate areas of research in the past, recent trend in adoption of OpenFlow has sparked research on combining the two. The benefits are very obvious, while SDN brings low maintenance and network complexity, distributed caching brings guaranteed low latency. Some recent findings in this includes \cite{Othman10} which proposes a application driven routing model on top of OpenFlow. Another notable work is \cite{Sakurauchi10} which proposes a system to dynamically redirect requests from a client based on content. Note that our approach is different from this, we have proposed a distributed caching system based on content based switching, thus extending \cite{Sakurauchi10}.

SDN has been considering L4-L7 extensions (say by Citrix or Qosmos), but those often take the form of DPI modules which do not integrate the routing decisions based upon the observed content. To operate on the granularity of content, many new architectures~\cite{NDN,Anand2011XIA,Paul2008Cacheandforward} have been proposed, which make identifying content and forwarding based upon data possible. However, these typically require to replace the whole network while our architectural model is based upon inserting a content-management layer in the SDN controller.

\section{Conclusion and future directions}
\label{sec:con}

In this paper, we have proposed and evaluated a generalization of the SDN philosophy to include non-forwarding elements into the network. This hybrid approach enables the end user to leverage the flexibility a traditional SDN provides coupled with the speedup that a traditional caching system provides. This work has a broad scope and impact, some immediate questions here are related to cache management: how does the caching policy impact the performance as observed by the end user? Can selfish caching help here? And can the set of primitives be expanded further beyond flows and contents, to include for instance some more complex workloads requiring synchronization of a wide set of network resources, including storage, compute and network? We plan to handle these problems as a natural extension to this work.



%

\bibliographystyle{ieeetr}
\bibliography{ContentNetPrimitive}

\end{document}